\documentclass[%
 reprint,
 amsmath,amssymb,
pra,
]{revtex4-1}

\usepackage{graphicx}
\usepackage{dcolumn}
\usepackage{bm}
\usepackage{float}
\usepackage{graphics,graphicx}
\usepackage{amsmath,mathrsfs,amsfonts,amssymb,amsthm}
\usepackage{color}
\usepackage{tabularx}
\usepackage{subfigure}
\usepackage{color}
\usepackage{fancyhdr}
\graphicspath{{../eps/}}
\DeclareGraphicsExtensions{.eps}

\def\v{\mathbf}
\def\mathscr{\mathcal}

\begin{document}

\preprint{APS/123-QED}

\title{Strongly Enhanced and Directionally Tunable Second-Harmonic Radiation by a Plasmonic Particle-in-Cavity Nanoantenna}
\author{Xiaoyan~Y.Z.~Xiong$^1$, Li~Jun~Jiang$^1$}
\email{ljiang@eee.hku.hk}
\author{Wei~E.I.~Sha$^1$}
\email{wsha@eee.hku.hk}
\author{Yat~Hei~Lo$^1$, Ming Fang$^{2,3}$, Weng Cho Chew$^{1,4}$, and Wallace C.H. Choy$^1$}
\affiliation{$^1$Department of EEE,~The University of Hong Kong, Hong Kong}%
\affiliation{$^2$Key Laboratory of Intelligent Computing and Signal Processing, Anhui University, Hefei, China}
\affiliation{$^3$Ames Laboratory and Department of Physics and Astronomy, Iowa State University, Iowa, USA}
\affiliation{$^4$Department of ECE, The University of Illinois at Urbana-Champaign, Illinois, USA}%

\begin{abstract}
Second-harmonic (SH) generation is tremendously important for nonlinear sensing, microscopy and communication system. One of the great challenges of current designs is to enhance the SH signal and simultaneously tune its radiation direction with a high directivity. In contrast to the linear plasmonic scattering dominated by a bulk dipolar mode, a complex surface-induced multipolar source at the doubled frequency sets a fundamental limit to control the SH radiation from metallic nanostructures. In this work, we harness plasmonic hybridization mechanism together with a special selection rule governing the SH radiation to achieve the high-intensity and tunable-direction emission by a metallic particle-in-cavity nanoantenna (PIC-NA). The nanoantenna is modelled with a first-principle, self-consistent boundary element method, which considers the depletion of pump waves. The giant SH enhancement arises from a hybridized gap plasmon resonance between the small particle and the large cavity that functions as a concentrator and reflector. Centrosymmetry breaking of the PIC-NA not only modifies the gap plasmon mode boosting the SH signal, but also redirects the SH wave with a unidirectional emission. The PIC-NA has a significantly larger SH conversion efficiency compared to existing literature. The main beam of the radiation pattern can be steered over a wide angle by tuning the particle's position.

\end{abstract}

\maketitle


\section{\label{sec:intro}Introduction}
Metallic nanostructures that support surface plasmon resonances have attracted significant attentions in nonlinear optics field due to their unique optical properties~\cite{ce01}. One such property is their ability to concentrate light energy in nanoscale volumes and subsequently boost the intensity of local fields near metal surfaces~\cite{ce02}. The plasmon-enhanced optical near fields allow weak nonlinear processes, which depend superlinearly on the local field, to be significantly enhanced leading to promising applications such as nonlinear optical sensing~\cite{ce03}, nonlinear optical characterization~\cite{ce06}, and nonlinear imaging~\cite{ce08}. Second-harmonic (SH) generation is a nonlinear process where two photons are converted into one photon at the doubled frequency. As an even-order nonlinear process, SH generation strongly depends on the symmetry of both the material being used and the structure being studied~\cite{ref02Optics}. SH generation is forbidden in the bulk of centrosymmetric media with the electric dipole approximation. However, the breaking of inversion symmetry at their surfaces results in the generation of SH waves~\cite{ref03Sipe}. In recent years, strong SH generation from a large variety of metallic nanostructures have been observed, including L-shaped~\cite{ce11} and G-shaped~\cite{ce13} particles, split-ring resonators~\cite{ce14}, dimers~\cite{ce15}, oligomers~\cite{ce16}, nanocups~\cite{ce17}, or even more complicated shapes~\cite{ce18}.

Efficient SH generation needs the presence of strong SH sources, i.e., strong nonlinear polarization currents at the nanostructure surface, as well as efficient outcoupling of SH signals into far-field. Several strategies have been developed to enhance SH generation from metallic nanostructures, including double resonances at both the fundamental and SH frequencies~\cite{ce19,ce20}, Fano resonances~\cite{ce21}, gap plasmon modes~\cite{ce22,ce28,Metasurfaces,Alu}, and symmetry breaking~\cite{ce26,ce27_1}. Due to a complex surface-induced multipolar source, the plasmon resonance, particle geometry, and local-field distribution simultaneously affect the SH process~\cite{ce29,ce30,ce31}. To facilitate nonlinear detection, sensing, microcopy and communication, not only a strong signal but also a high directional emission is essential. This poses a great challenge to current designs of nonlinear optical devices. Although the directional control of SH radiation from non-centrosymmetric materials has been reported recently~\cite{AlGaAs}, its physical principles and designs in centrosymmetric metal structures are still unexplored.

In this work, a metallic particle-in-cavity nanoantenna (PIC-NA) as a prototype is proposed to overcome the challenge. To the best of our knowledge, it is the first time that both a giant SH enhancement and a directionally tunable SH radiation have been simultaneously realized. The PIC-NA comprises a small nanosphere inside a large nanocup cavity. Although its linear responses were investigated~\cite{ce32,ce33}, SH radiation from the structure shows extraordinary features and has not been studied and understood yet.

\vfill

\section{\label{sec:methods}Method}
SH generation can be described by the two coupled linear wave equations
\begin{eqnarray}\label{eq1}
\nabla^2\v{E}^{(\omega)}+{k^{(\omega)}}^2\v{E}^{(\omega)}=-\frac{\omega^2}{\varepsilon_0c^2}\v{P}^{(\omega),\textrm{NL}}
-\frac{i\omega}{\varepsilon_0c^2}\v{J}^{(\omega)}_{\textrm{pump}}\\
\nabla^2\v{E}^{(\Omega)}+{k^{(\Omega)}}^2\v{E}^{(\Omega)}=-\frac{\Omega^2}{\varepsilon_0c^2}\v{P}^{(\Omega),\textrm{NL}}
\end{eqnarray}
where $\v{J}^{(\omega)}_{\textrm{pump}}$ is a pump source producing the incident fundamental field. The nonlinear polarizations $\v{P}^{(\omega),\textrm{NL}}=\varepsilon_0\overline{\overline\chi}_{sd}^{(2)}:\v{E}^{(\Omega)}\v{E}^{(\omega)*}$ and $\v{P}^{(\Omega),\textrm{NL}}=\varepsilon_0\overline{\overline\chi}_{su}^{(2)}:\v{E}^{(\omega)}\v{E}^{(\omega)}$ are the nonlinear source terms for the fundamental and SH fields, where $\Omega=2\omega$. Here, $\overline{\overline\chi}_{sd}^{(2)}$ and $\overline{\overline\chi}_{su}^{(2)}$ are the surface second-order nonlinear susceptibility tensors related to the down-conversion and up-conversion processes, respectively. And $\v{E}^{(\omega)}$ and $\v{E}^{(\Omega)}$ denote the fundamental and SH electric fields, respectively. $k^{(\omega)}$  and $k^{(\Omega)}$ are the corresponding wave numbers. To fully capture the pump depletion effect and cross couplings between the fundamental and SH fields, Equations (1) and (2) are solved self-consistently by a first-principle boundary element method with an initial condition $\v{P}^{(\omega),\textrm{NL}}=0$. The boundary element method can model the SH scattering from arbitrarily shaped structures efficiently since it only requires surface discretization. In addition, experimentally tabulated material parameters can be used directly.

The computational domain of the electromagnetic field is divided into the interior of the metal and the exterior medium separated by the interface $S$. The object is illuminated by the plane wave source. By invoking the Love's equivalence principle, the equivalent currents on the external side of $S$ produce the scattered field in the exterior region and null field in the interior region while the equivalent currents on the internal side of $S$ produce the total field in the interior region and null field in the exterior region.
\begin{eqnarray}\label{ChSHGeq1}
\!\!\!\!\!\!\left. \begin{array}{ll}
\v{r}\in V_\ell, \, \v{E}_\ell^{(\nu)}(\v{r})\\
\v{r}\notin V_\ell,\,  0
\end{array} \right\}
=i\nu\mu_\ell\int_S\overline{\v{G}}^{(\nu)}\left(\mathbf{r},\mathbf{r}^\prime\right)\cdot\v{J}_\ell^{(\nu)}(\v{r}^\prime)d\v{r}^\prime \\ \nonumber
-\int_S\nabla\times\overline{\v{G}}^{(\nu)}\left(\mathbf{r},\mathbf{r}^\prime\right)\cdot\v{M}_\ell^{(\nu)}(\v{r}^\prime)d\v{r}^\prime
\end{eqnarray}
\begin{eqnarray}\label{ChSHGeq2}
\!\!\!\!\!\!\left. \begin{array}{ll}
\v{r}\in V_\ell, \, \v{H}_\ell^{(\nu)}(\v{r})\\
\v{r}\notin V_\ell,\,  0
\end{array} \right\}
=i\nu\epsilon_\ell\int_S\overline{\v{G}}^{(\nu)}\left(\mathbf{r},\mathbf{r}^\prime\right)\cdot\v{M}_\ell^{(\nu)}(\v{r}^\prime)d\v{r}^\prime \\ \nonumber
+\int_S\nabla\times\overline{\v{G}}^{(\nu)}\left(\mathbf{r},\mathbf{r}^\prime\right)\cdot\v{J}_\ell^{(\nu)}(\v{r}^\prime)d\v{r}^\prime
\end{eqnarray}
where $\overline{\v{G}}^{(\nu)}(\v{r},\v{r}^\prime)=(\overline{\v{I}}+{k^{(\nu)}}^{-2}\nabla\nabla)\exp{(ik^{(\nu)}R)}/4\pi R$ is the dyadic Green's function at the frequency $\nu$ ($\nu=\omega,\,\Omega$ denote the fundamental and SH frequencies, respectively). $R=|\mathbf{r}-\mathbf{r}^\prime|$. Here, $\ell=e,\,i$ denote the exterior and interior regions of the object, respectively. $\v{J}_\ell^{(\nu)}$ and $\v{M}_\ell^{(\nu)}$ are the equivalent electric and magnetic currents placed on the outer ($\ell=e$) and inner ($\ell=i$) sides of the surface of a nanostructure. The discontinuities of electromagnetic fields at the interface are due to the existence of current sources $\v{J}_0^{(\nu)}$ and $\v{M}_0^{(\nu)}$ at the boundary.
\begin{eqnarray}\label{eq4}
\left\{ \begin{array}{ll}
\v{n}\times\left(\v{E}_i^{(\nu)}-\v{E}_e^{(\nu)}  \right)=-\v{M}_0^{(\nu)}\\
\v{n}\times\left(\v{H}_i^{(\nu)}-\v{H}_e^{(\nu)}  \right)=\v{J}_0^{(\nu)}
\end{array} \right.
\end{eqnarray}
where $\v{n}$ is the outer normal vector of the surface. The equivalent currents satisfy
\begin{eqnarray}\label{eq7}
\left\{ \begin{array}{cc}
\v{J}_i^{(\nu)}+\v{J}_e^{(\nu)}=\v{J}_0^{(\nu)} \\
\v{M}_i^{(\nu)}+\v{M}_e^{(\nu)}=\v{M}_0^{(\nu)}
\end{array} \right.
\end{eqnarray}

The surfaces of the nanostructures are discretized with triangular mesh. The equivalent currents are expanded as Rao-Wilton-Glisson (RWG) basis functions~\cite{ce41}. A matrix system is then constructed by exploiting the Galerkin testing procedure~\cite{ceGalerkin}. A modified Poggio-Miller-Chang-Harrington-Wu-Tsai (PMCHWT) formulation~\cite{ref06Kauranen,ce14_2} is used to ensure accurate solutions even at resonant conditions. The PMCHWT matrix equation can be written as
\begin{eqnarray}\label{eq9}
C^{(\nu)}x^{(\nu)}=y^{(\nu)},
\end{eqnarray}
with the impedance matrix
\begin{eqnarray}\label{eq10}
C^{(\nu)}= \left [\begin{array}{cccc}
\mathscr{L}_e^{(\nu)}& \mathscr{K}_e^{'(\nu)} & -\mathscr{L}_i^{(\nu)} & -\mathscr{K}_i^{'(\nu)} \\
-\mathscr{K}_e^{'(\nu)}& \eta_e^{-2}\mathscr{L}_e^{(\nu)} & \mathscr{K}_i^{'(\nu)} & -\eta_i^{-2}\mathscr{L}_i^{(\nu)} \\
\mathscr{I} & 0 & \mathscr{I} & 0 \\
0 &\mathscr{I} & 0 &  \mathscr{I} \nonumber \\
\end{array}\right ]
\end{eqnarray}
where $\mathscr{L}_\ell^{(\nu)}=i\nu\mu_\ell\overline{\v{G}}^{(\nu)}$ and $\mathscr{K}_\ell^{'(\nu)}$ is the principal value part of the $\mathscr{K}_\ell^{(\nu)}$ operator with $\mathscr{K}_\ell^{(\nu)}=-\nabla\times\overline{\v{G}}^{(\nu)}$. $\eta_\ell=\sqrt{\mu_\ell/\epsilon_\ell},$ with $\ell=e,\,i$.
The vector of unknowns $x^{(\nu)}$ and the excitation $y^{(\nu)}$ are
\begin{eqnarray}\label{eq11}
x^{(\nu)}= \left [\begin{array}{c}
\v{J}_e^{(\nu)} \\
\v{M}_e^{(\nu)} \\
\v{J}_i^{(\nu)} \\
\v{M}_i^{(\nu)} \\
\end{array}\right ], \qquad
y^{(\nu)}= \left [\begin{array}{c}
\frac{1}{2}\v{n}\times\v{M}_0^{(\nu)} \\
-\frac{1}{2}\v{n}\times\v{J}_0^{(\nu)} \\
\v{J}_0^{(\nu)} \\
\v{M}_0^{(\nu)} \nonumber \\
\end{array}\right ]
\end{eqnarray}

\noindent According to the boundary conditions~\cite{HeinzBookChapter}, we have
\begin{eqnarray}\label{ChSHGeq31}
\qquad\left\{ \begin{array}{ll}
\v{J}_0^{(\omega)}=\v{n}\times\v{H}^{(\omega, inc)} -i\omega\v{P}_{St}^{(\omega)}\\
\v{M}_0^{(\omega)}=-\v{n}\times\v{E}^{(\omega, inc)}+ \frac{1}{\epsilon^\prime}\v{n}\times\nabla_S P_{Sn}^{(\omega)}
\end{array} \right.
\end{eqnarray}
\begin{eqnarray}\label{ChSHGeq34}
\!\!\!\!\!\!\!\!\!\!\!\!\!\!\!\!\!\!\!\!\!\!\!\!\!\!\left\{ \begin{array}{ll}
\v{J}_0^{(\Omega)} = -i\Omega\v{P}_{St}^{(\Omega)} \\
\v{M}_0^{(\Omega)}= \frac{1}{\epsilon^\prime}\v{n}\times\nabla_S P_{Sn}^{(\Omega)} \\
\end{array} \right.
\end{eqnarray}
where $\v{P}_{St}^{(\nu)}=-\v{n}\times\v{n}\times\v{P}_S^{(\nu)}$ and $P_{Sn}^{(\nu)}=\v{n}\cdot\v{P}_S^{(\nu)}$ are the tangential and normal components of the nonlinear polarization with $\v{P}_S^{(\omega)}=\epsilon_0\overline{\overline\chi}_{s}^{(2)}:\v{E}^{(\Omega)}\v{E}^{(\omega)*}$  and $\v{P}_S^{(\Omega)}=\epsilon_0\overline{\overline\chi}_{s}^{(2)}:\v{E}^{(\omega)}\v{E}^{(\omega)}$. Here, $\epsilon^\prime$ is the selvedge region permittivity.

Equation \eqref{eq9} can be used to solve both the fundamental and SH fields by setting $\nu=\omega\,,\Omega$, respectively. As indicated in \eqref{ChSHGeq31}, the source terms of the fundamental field include the incident plane wave and the nonlinear polarization source coupled back from the SH field. The second-order susceptibility acts as the coupling coefficient. The coupling between the fundamental and SH fields is modeled by solving these two processes iteratively, with the updated source terms given in the previous iteration. In the numerical implementation procedure, we initially calculate the equivalent fundamental electric and magnetic currents induced by the incident plane wave by setting the nonlinear polarization $\v{P}_S^{(\omega)}$ to be zero. In order to compute the initial value of the SH nonlinear polarization, we need to evaluate the fundamental field on the internal side of the surface $S$. It is given by the equivalent surface currents directly:
\begin{eqnarray}\label{eq15}
\v{E}_{it}^{(\omega)}\mid_{S^{-}}= -\v{n}\times\v{M}_i^{(\omega)} \\
E_{in}^{(\omega)}\mid_{S^{-}}= i\frac{\nabla_S\cdot\v{J}_i^{(\omega)}}{\omega\epsilon}
\end{eqnarray}
With this obtained fundamental field, the initial value of the SH nonlinear polarization is first  calculated, from which the initial value of the SH field can be solved. After these initial values being obtained, we can iteratively solve for the field unknowns at the fundamental and SH frequencies to arrive at the self-consistent solution.

In this work, the dielectric constant for gold is taken from experimental data~\cite{ref10Kauranen}. The surface susceptibility of gold has three independent components, $\overline{\overline\chi}_{s,\perp\perp\perp}^{(2)}=1.59\times10^{-18}\,\textrm{m}^2/V$, $\overline{\overline\chi}_{s,\parallel\parallel\perp}^{(2)}=\overline{\overline\chi}_{s,\parallel\perp\parallel}^{(2)}=4.63\times10^{-20}\,\textrm{m}^2/V$, and $\overline{\overline\chi}_{s,\perp\parallel\parallel}^{(2)}=0$~\cite{ceKrause}. Here, $\perp$ and $\parallel$ denote the component normal and tangential to the surface, respectively.

\section{\label{sec:results}Results}
\subsection{Schematic Configuration}
\begin{figure}[!h]
\begin{center}
\includegraphics[width=3.1in]{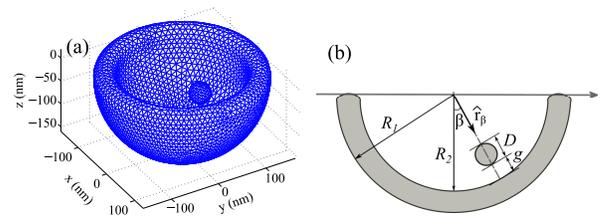}
  \caption{Schematic of the particle-in-cavity nanoantenna (PIC-NA). (a) Triangular mesh with $7,516$ triangles used for the theoretical model; (b) The cross section ($yoz$ plane) of the PIC-NA. A nanosphere (diameter $D$) is inside a nanocup cavity (external and inner radii being $R_1$ and $R_2$) separated by a gap $g$. The location of the sphere can be identified by a position vector $\hat{r}_{\beta}$, which connects the sphere center and origin of coordinates. The twist angle between the position vector and $z$ axis is $\beta$. Centrosymmetry of the structure is broken when $\beta\neq0$. The PIC-NA is illuminated by a $y$-polarized plane wave at a normal incidence from the top.}
\label{fig1}
\end{center}
\end{figure}

The configuration of the investigated PIC-NA is depicted in Fig.~\ref{fig1}. A gold nanosphere with a diameter of $D$ is placed inside a gold rounded-edge nanocup cavity separated by a small gap $g$. The nanocup cavity is a truncated hemispherical nanoshell with the external and inner radii being $R_1$ and $R_2$, respectively. Depending on the twist angle between the nanosphere and nanocup (as shown in Fig.~\ref{fig1}(b)), the nanoantenna has symmetric ($\beta=0$) or asymmetric ($\beta\neq0$) geometry. The PIC-NA is illuminated by a $y$-polarized plane wave at a normal incidence from the top. The experimental method with the detailed fabrication processes for the PIC-NA has been presented in~\cite{ce32}. The linear and nonlinear scattering from the PIC-NA are modelled by the developed boundary element method. The mesh configuration is shown in Fig.~\ref{fig1}(a).

Placing the metallic nanoparticle inside the nanocup cavity produces extremely strong field enhancements at the particle-cavity gap especially when one of the cavity modes is hybridized with the cavity-dressed nanoparticle mode. The dimensions of the cavity are optimized to be the external radius $160\,\textrm{nm}$ and the inner radius $120\,\textrm{nm}$, so that strong enhancements occur at the visible light regime. To simplify the analysis, the gap is fixed to $5\,\textrm{nm}$ and the PIC structure is assumed to be embedded in air background.

\begin{figure}[!h]
\begin{center}
\includegraphics[width=3.2in]{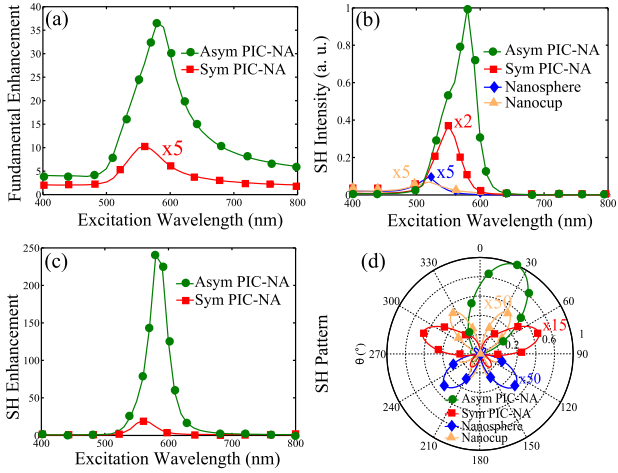}
  \caption{Linear and SH responses of the PIC-NA with a $40\,\textrm{nm}$ gold nanosphere and $\beta=30^\circ$. (a) Fundamental enhancement spectra. (b) Normalized SH intensity spectra of four different nanostructures: single nanosphere with the radius of $40\,\textrm{nm}$, single nanocup with the external and inner radii of $160\,\textrm{nm}$ and $120\,\textrm{nm}$, symmetric PIC-NA, and asymmetric PIC-NA. The SH intensities are normalized by the maximum intensity obtained by the asymmetric PIC-NA. (c) SH enhancement spectra. (d) SH radiation patterns ($yoz$ plane) of the four nanostructures at their resonant wavelengths.}
\label{fig2}
\end{center}
\end{figure}

\subsection{Giant Enhancement of SH Signals}
\begin{figure}[!h]
\begin{center}
\includegraphics[width=3.2in]{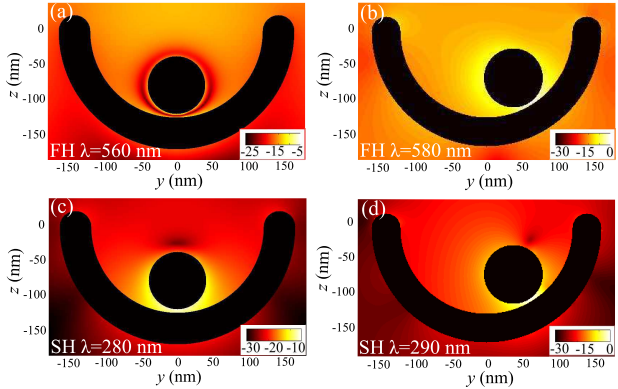}
  \caption{Normalized fundamental and SH scattered fields evaluated at the $yoz$ plane for the symmetric and asymmetric PIC-NA with a $40\,\textrm{nm}$ gold nanosphere. (a, b) fundamental field and (c, d) SH field (all shown in a logarithmic scale). (a, c) symmetric PIC-NA at the resonant wavelength of $\lambda=560\,\textrm{nm}$; (b, d) asymmetric PIC-NA ($\beta=30^\circ$) at the resonant wavelength of $\lambda=580\,\textrm{nm}$. The amplitude of the fundamental field of the symmetric PIC-NA is normalized by the maximum amplitude of the fundamental field of the asymmetric one. So does the SH field.}
\label{fig3}
\end{center}
\end{figure}

The fundamental field is investigated at a fixed point in the gap center. The field within the gap is almost uniform when the nanosphere is close to the nanocavity surface. Hence, the enhancement factor can be defined as the ratio of the magnitude of scattered field to that of the incident field at the fixed point. Figure~\ref{fig2}(a) shows the fundamental enhancement spectra of the PIC-NA with a $40\,\textrm{nm}$ gold nanosphere. The peak enhancements of roughly $2$ at $\lambda=560\,\textrm{nm}$ and $37$ at $\lambda=580\,\textrm{nm}$ are found for the symmetric and asymmetric PIC-NAs, respectively. This indicates that the asymmetric configuration strongly modifies the cavity-particle mode hybridization. Due to the centrosymmetry breaking with $\beta\neq0$, the polarized incident electric field is not orthogonal (vertical) to the position vector of sphere ($\hat{r}_{\beta}$). Therefore, compared to the fundamental scattered field from the symmetric PIC-NA in Fig.~\ref{fig3}(a), a stronger gap plasmon resonance occurs in the asymmetric design. (see Fig.~\ref{fig3}(b)). The giant near-field enhancement at the fundamental frequency allows for an efficient SH generation.

\begin{figure}[!h]
\begin{center}
\includegraphics[width=2.5in]{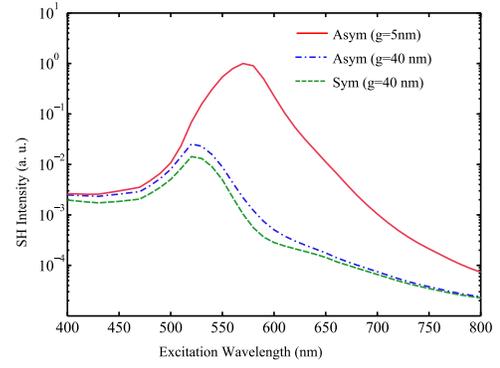}
  \caption{Normalized SH intensity spectra of the PIC-NA with different gap sizes and symmetries. The radius of the gold sphere is fixed to $40\,\textrm{nm}$.}
\label{fig4}
\end{center}
\end{figure}

To understand the SH enhancement by the PIC-NA, the SH intensities (total scattered power) of four different structures are comparatively studied (see Fig.~\ref{fig2}(b)). Thanks to the mode hybridization, the two coupled PIC-NAs have much stronger SH intensities than both the single nanosphere and the single nanocup. Furthermore, the asymmetric PIC-NA undoubtedly gains the strongest SH signal. The SH enhancement factor, defined as the ratio of the SH scattered power of the PIC-NA to the summation of those of the nanosphere and nanocup, is depicted in Fig.~\ref{fig2}(c). Intensity enhancements as high as $240$ and $20$ can be obtained for the asymmetric and symmetric PIC-NAs at the resonant wavelengths of $\lambda=580\,\textrm{nm}$ and $\lambda=560\,\textrm{nm}$, respectively. The correlation between the fundamental enhancement (Fig.~\ref{fig2}(a)) and the SH intensity spectra (Fig.~\ref{fig2}(c)) suggests that the enhanced SH field is attributed to the nanocup concentrator that focuses large optical energy into the small gap. The metallic nanocup cavity resonantly concentrates light inside the cavity. Meanwhile, the enhanced field excites the plasmonic mode of the adjacent small nanosphere forming the gap plasmon resonance. Consequently, the light is predominantly confined at the tiny nanogap. From comparing Figs.~\ref{fig3}(d) to (c), asymmetric PIC-NA has a pronounced SH field confinement at the nanogap.

There are two physical reasons for the SH enhancement. One is the breaking of the overall centrosymmetry of the PIC-NA with the twist angle $\beta\neq 0$. The other is originated from gap plasmon resonance. To understand the roles of centrosymmetry breaking and gap plasmon resonance in boosting the SH signal, we calculate the SH intensities for the symmetric PIC-NA with a large gap size ($g$=40 nm), the asymmetric one with a large gap size ($g$=40 nm), and the asymmetric one with a small gap size ($g$=5 nm). From the results in Fig.~\ref{fig4}, gap plasmon resonance is a dominating factor for the SH enhancement.

\subsection{Beam Steering for SH Radiation}
Besides a remarkable SH enhancement, the SH radiation patterns of the PIC-NA presented in Fig.~\ref{fig2}(d) show extraordinary features, which cannot be achieved by other metallic structures investigated in literature. According to a special selection rule~\cite{ce14_2,cesw02} governing the SH radiation, SH radiation from a metallic object is strictly zero along the incident $z$ direction, if the projection of the object onto the $xoy$ plane is centrosymmetric. The special selection rule coincides with the parity conservation law~\cite{cesw03}. The net nonlinear polarization is nonzero along the $z$ direction due to the phase retardation from the fundamental incident wave propagating along the $-z$ direction. For the nanosphere, nanocup and symmetric PIC-NA with centrosymmetric geometries at the $xoy$ plane, they all show null radiations along the $\pm z$ direction and symmetric radiations at the $xoy$ plane with two main beams. Different from the nanosphere with a forward SH radiation, the nanocup, which has a significant backward SH scattering, behaves like a big reflector for the SH generation. Intuitively, the almost-lateral radiation of the symmetric PIC-NA is an enhanced superposition of the radiations from the nanosphere and nanocup. Amazingly, a unidirectional SH radiation is observed for the asymmetric PIC-NA (see Fig.~\ref{fig2}(d)).

\begin{figure}[!h]
\begin{center}
\includegraphics[width=3.3in]{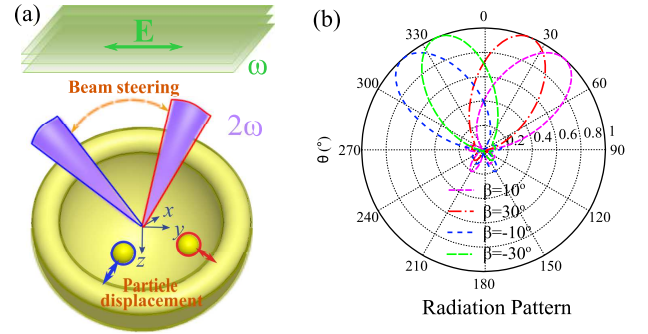}
  \caption{(a) Conceptual realization of SH beam steering by manipulating particle displacement. The SH radiation direction is approximately vertical to the dominated nonlinear polarization direction along the position vector of the sphere ($\hat{r}_{\beta}$ in Fig.~\ref{fig1}), which is denoted with the double headed arrow. The dominated nonlinear polarization is due to the optical field enhancement by the gap plasmon mode. (b) The dependence of SH radiation patterns on the twist angle $\beta$ illustrated in Fig.~\ref{fig1}.}
\label{fig5}
\end{center}
\end{figure}

Moreover, as illustrated in Fig.~\ref{fig5}, one can tune the emission direction over a wide angle by placing the nanosphere at different positions. This fantastic beam-steering feature greatly facilitates the far-field detection of generated SH waves. The $y$ polarized incident electric field can be decomposed as the two polarized field components parallel and perpendicular to the position vector of the sphere ($\hat{r}_{\beta}$ in Fig.~\ref{fig1}), respectively. The parallel component is responsible for the gap plasmon coupling generating a dominant nonlinear polarization (See Fig.~\ref{fig5}(a)). While the perpendicular component generates a weak nonlinear polarization. As the twist angle $|\beta|$ increases, with the aid of the cavity reflector, the dominant parallel nonlinear source induces the unidirectional SH radiation whose direction is closer to 0 degree at the far-field zone (See Fig.~\ref{fig5}(b)). The extraordinary beam-steering feature of the SH radiation is also confirmed by Maxwell-hydrodynamic model as illustrated in Fig.~\ref{figS-MH}. The simulation results by the Maxwell-hydrodynamic model agree with the boundary element method results well. Regarding details of Maxwell-hydrodynamic model, please see Appendix A.

\begin{figure}[!h]
\begin{center}
\includegraphics[width=3.2in]{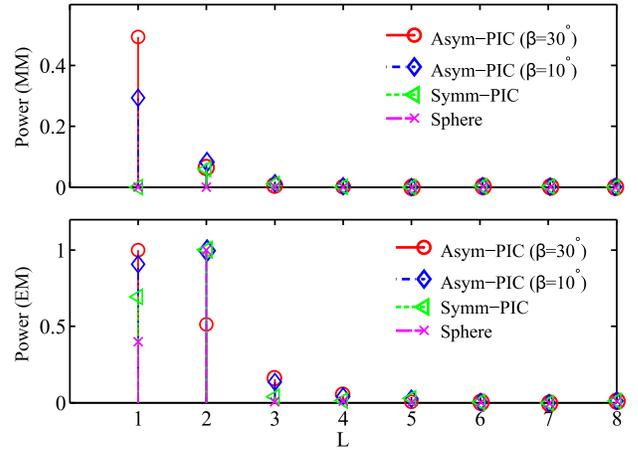}
  \caption{Multipole expansion of four different nanostructures at the SH frequency: single nanosphere with the radius of $40\,\textrm{nm}$, symmetric PIC-NA, asymmetric PIC-NA with $\beta=10^\circ$, and asymmetric PIC-NA with $\beta=30^\circ$. MM and EM denote magnetic multipole and electric multipole, respectively.}
\label{fig7}
\end{center}
\end{figure}

\begin{figure}[!h]
\begin{center}
\includegraphics[width=3.2in]{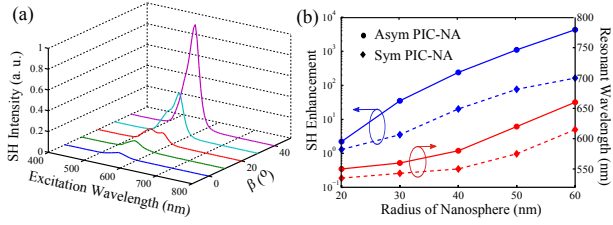}
  \caption{(a) Normalized SH intensity spectra of the PIC-NA with a $40\,\textrm{nm}$ nanosphere located at different positions ($\beta$ varies from $0^\circ$ to $40^\circ$). (b) SH enhancement factors and corresponding resonant wavelengths of the PIC-NAs with the radius of the nanosphere varying from $20$ to $60$ nm.}
\label{fig6}
\end{center}
\end{figure}

Next, we analyze the multipole interaction in the PIC-NA, which is essential to the SH radiation (See Fig.~\ref{fig7}). Regarding a single sphere, only a weak electric dipole and a strong electric quadrupole exist at the plasmonic resonance. The multipole expansion result agrees with the analyses in~\cite{HeinzPRL,deBeerPRB} well. Due to a weak plasmon coupling in the symmetric PIC-NA, the ignorable magnetic quadrupole field and slightly enhanced electric dipole field are observed. For the centrosymmetric sphere and PIC-NA, closed loop currents at the SH frequency cannot be supported and thus magnetic dipole is exactly zero. However, the asymmetric PIC-NA exhibits a significant amplification for both magnetic dipole and electric dipole fields. The loop electric and magnetic currents are induced by breaking the centrosymmetry and by the strong gap plasmon coupling. Additionally, a small (high-order) electric octupole field is excited as well. Consequently, the constructive and destructive interference between the magnetic dipole, electric dipole, and electric quadrupole fields are responsible for the unidirectional SH radiation.

\subsection{Universal Gap Plasmon Mode}
To further understand the role of the particle-cavity mode hybridization in affecting the SH response, we calculate the SH spectra of the PIC-NA when the twist angle $\beta$ varies from $0^\circ$ to $40^\circ$ as displayed in Fig.~\ref{fig6}(a). As the twist angle becomes large, the increasing alignment between the polarization direction of the incident electric field ($y$ polarized) and the position vector of the nanosphere ($\hat{r}_{\beta}$ in Fig.~\ref{fig1}) induces a stronger gap plasmon resonance and thus a stronger nonlinear polarization. Consequently, the PIC-NA generates SH waves with increasing intensity. Additionally, Fig.~\ref{fig6}(b) shows the boosted SH generation for both the symmetric and asymmetric PIC-NAs, when the nanosphere radius varies from $20$ to $60$ nm. The enlarged coupling surface between the nanosphere and nanocup leads to the enhancement. It is known that larger nanosphere has a longer resonant wavelength. Similarly, for the particle-cavity hybridized system, the resonant wavelengths are red shifted as the nanosphere radius grows, which is also shown in Fig.~\ref{fig6}(b). Here, the SH enhancement factor as high as $10^4$ is obtained with a 60 nm nanosphere.

\begin{figure}[!htb]
\begin{center}
\includegraphics[width=3.5in]{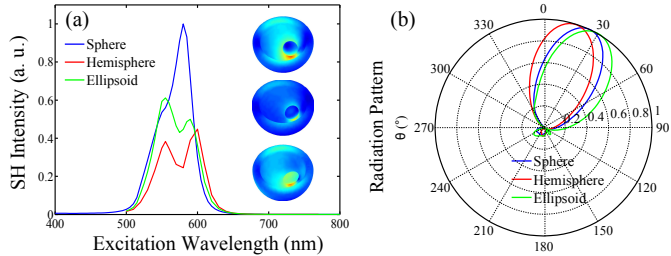}
  \caption{(a) SH intensity of the asymmetric PIC-NA with different particle shapes: sphere (radius $r=40\,\textrm{nm}$), hemisphere (radius $r=40\,\textrm{nm}$) and ellipsoid (major radius $r=40\,\textrm{nm}$, axial ratio=1.5). Other geometric parameters are: $R_1=160\,\textrm{nm}$, $R_2=120\,\textrm{nm}$, $g=5\,\textrm{nm}$, and $\beta=30^\circ$.  The inset in (a) shows the fundamental electric current distributions of the PIC-NA with different particle shapes. (b) SH radiation pattern  (E-plane) of the asymmetric PIC-NA with different particle shapes.}
\label{figS4}
\end{center}
\end{figure}

The directional SH radiation has been achieved with different shaped nanoparticles in the cavity involving sphere, hemisphere, and ellipsoid as shown in Fig.~\ref{figS4}. The centrosymmetry breaking induced gap plasmon mode (See the inset in Fig.~\ref{figS4}(a)), is a universal physical origin of the unidirectional SH emission with a high directivity as depicted in Fig.~\ref{figS4}(b).

\subsection{Comparisons to Designs in Literature}
\begin{figure}[!htb]
\begin{center}
\includegraphics[width=3.5in]{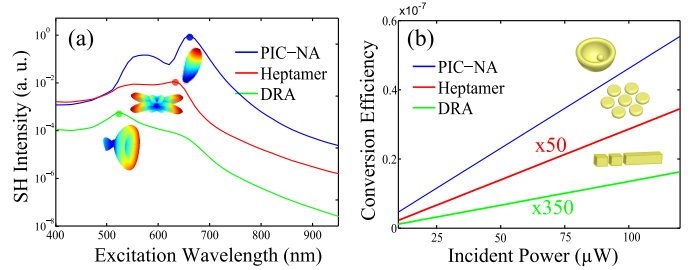}
  \caption{SH intensities (a) and conversion efficiencies (b) of the asymmetric PIC-NA, heptamer and double resonant antenna (DRA), all with the same material of gold and the same surface area. The parameters of the PIC-NA are: $R_1=160\,\textrm{nm}$, $R_2=120\,\textrm{nm}$, $D=120\,\textrm{nm}$, $g=5\,\textrm{nm}$ and $\beta=40^\circ$. The SH conversion efficiency is defined as the SH scattered power over the product of incident flux and geometric cross-section area of a structure.}
\label{fig9}
\end{center}
\end{figure}

To demonstrate a significant breakthrough in our design, the asymmetric PIC-NA is compared to other representative designs in literature (see Fig.~\ref{fig9}). To carry out a fair comparison, geometries of the heptamer~\cite{ce16} and double resonant antenna~\cite{ce19} are scaled to achieve the same surface area as the PIC-NA. The PIC-NA has a significantly larger SH conversion efficiency simultaneously with a high directivity.

\subsection{Self-Consistent Solution versus Undepleted Pump Approximation}
\begin{figure}[!h]
\begin{center}
\includegraphics[width=2.0in]{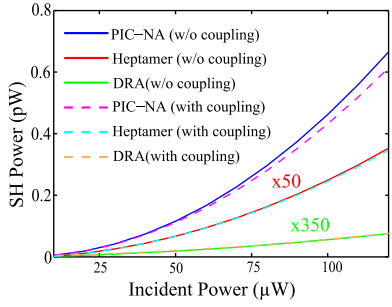}
  \caption{SH intensities of the asymmetric PIC-NA, heptamer and double resonant antenna (DRA) as a function of the incident power. Solid lines: calculated without the back coupling effect; dashed lines: calculated with the back coupling effect.}
\label{figS1}
\end{center}
\end{figure}

\begin{figure}[!htb]
\begin{center}
\includegraphics[width=3.5in]{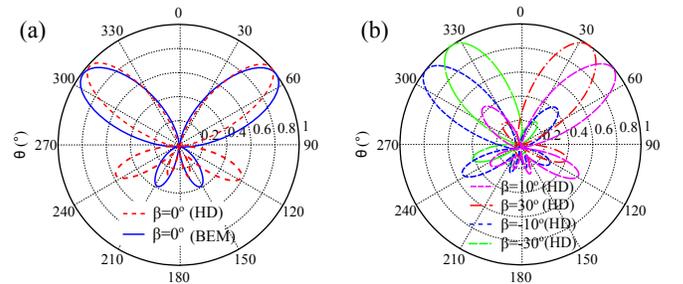}
  \caption{(a) SH radiation patterns (E-plane) of the PIC-NA calculated by the boundary element method (BEM) and hydrodynamic (HD) model ($\beta=0^\circ$). (b) The dependence of SH radiation patterns (E-plane) on the twist angle $\beta$ calculated by the hydrodynamic (HD) model. The parameters for the hydrodynamic model are: the phenomenological damping frequency $\gamma=1.075\times10^{14}$ rad/s, electron density $n_0=5.98\times10^{28}\,/\textrm{m}^{3}$ and incident wavelength $\lambda=520\,\textrm{nm}$.}
\label{figS-MH}
\end{center}
\end{figure}

Additionally, the giant SH enhancement breaks undepleted pump approximation under a high-power laser illumination as shown in Fig.~\ref{figS1}. The SH intensities of the asymmetric PIC-NA, heptamer~\cite{ce16} and double resonant antenna (DRA)~\cite{ce19} are calculated as a function of the incident power, which is defined as the product of incident flux and geometric cross-section area of a structure. For non-self-consistent method with an undepleted pump approximation~\cite{ref06Kauranen}, the back coupling from the SH field to fundamental field is ignored. For the proposed self-consistent method, the back coupling is fully captured. Due to a giant SH enhancement achieved by the asymmetric PIC-NA, a significant deviation between the two methods can be observed under a high-power laser illumination. Although such a high-power irradiation may melt the gold particle, our method provides a novel way to evaluate the design of SH enhancement and to analyze a potentially significant pump depletion in periodic plasmonic nanostructures, where a long-distance light-matter nonlinear interaction will occur at the periodic directions.

\section{Conclusion}
In conclusion, SH generation from a 3D plasmonic PIC-NA has been theoretically investigated. The PIC-NA achieves a strong SH enhancement and a tunable directional radiation simultaneously. By manipulating the displacement of nanosphere, the symmetry dependent mode hybridization between the nanosphere and nanocap forms a tunable gap plasmon mode, which results in the giant SH enhancement and the tunable beam steering. The class of PIC-NAs could offer highly controllable SH radiation by changing the size, position and material of the nanoparticle. Our work is fundamentally and practically important to nonlinear sensing, detection, spectroscopy, and on-chip communication.

\section{Acknowledgements}
This work was supported by the Research Grants Council of Hong Kong (GRF 716713, GRF 17207114, and GRF 17210815), National Science Foundation of China (Nos. 61271158 and 61201122), Hong Kong ITP/045/14LP, Hong Kong UGC AoE/PC04/08, the Collaborative Research Fund (No. C7045-14E) from the Research Grants Council of Hong Kong, and Grant CAS14601 from CAS-Croucher Funding Scheme for Joint Laboratories.

\section{APPENDIX A}\label{Secappend}
To model SH generation from the PIC-NA, Maxwell-hydrodynamic equations are solved nonperturbatively and self-consistently by the explicit FDTD method with Yee grids \cite{Taflove,Hydrodynamic1,Hydrodynamic2}. Using the pure scattered field technique, the governing equations including Maxwell, hydrodynamic, and current continuity equations are given by
\begin{equation}\label{eqmh1}
\nabla\times \v{E}_{sca} =  - \mu _0\frac{\partial \v{H}_{sca}} {\partial t}, \quad
\nabla\times \v{H}_{sca} = \v{J} + \epsilon_0\frac{\partial \v{E}_{sca}}{\partial t}
\end{equation}

\begin{equation}\label{eqmh2}
\frac{\partial \v{v}}{\partial t} + \v{v} \cdot \nabla \v{v} + \gamma \v{v} =
 \frac{-e}{m}( \v{E}_{tot}  +
 \mu _0\v{v} \times \v{H}_{tot}   ) + \nabla V_P
\end{equation}

\begin{equation}\label{eqmh3}
\frac{\partial n}{\partial t} =  - \nabla \cdot (n\v{v}),\quad{\v{J}} =  - en\v{v}
\end{equation}

\noindent where $e$ and $m$ are the elementary charge and electron mass. $\mathbf{J}$ is the polarization current, $\mathbf{v}$ is the electron velocity, and $n$ is the electron density. ${\mathbf{E}}_{sca}$ and ${\mathbf{H}}_{sca}$ are the scattered electric and magnetic fields produced by the polarization current. ${\mathbf{E}}_{tot}$ and ${\mathbf{H}}_{tot}$ are the total electric and magnetic fields, which are the summations of incident pump (laser source or plane wave source) fields and scattered fields. ${\mathbf{v}} \cdot \nabla {\mathbf{v}}$ is the convective acceleration term and $\gamma$ is the scattering rate (related to the metallic loss). The potential $V_P$ including Hartree, exchange, Fermi, Bohm potentials can be incorporated to capture the nonlocal and quantum effects. Here, the current continuity equation \eqref{eqmh3} is employed to connect Maxwell equation \eqref{eqmh1} and hydrodynamic equation \eqref{eqmh2}.

\end{document}